\documentclass[conference]{IEEEtran}
\usepackage[utf8]{inputenc}
\usepackage{graphicx}
\usepackage{hyperref}
\usepackage{amsmath, amssymb, amsfonts}
\usepackage{booktabs}
\usepackage{listings}
\usepackage[table,xcdraw]{xcolor}
\usepackage{array}
\usepackage{caption}
\usepackage{float}
\usepackage{colortbl}
\usepackage{subcaption} 

\renewcommand{\arraystretch}{1.5}
\title{Combining Threat Intelligence with IoT Scanning to Predict Cyber Attacks}

\author{
\IEEEauthorblockN{Jubin Abhishek Soni}
\IEEEauthorblockA{\textit{Researcher} \\
Senior Member, IEEE \\
San Francisco, CA, USA \\
jubin.soni@ieee.org}
\and
\IEEEauthorblockN{Amit Anand}
\IEEEauthorblockA{\textit{Researcher} \\
Senior Member, IEEE \\
Austin, TX, USA \\
amit.anand@ieee.org}
\and
\IEEEauthorblockN{Rajesh Kumar Pandey}
\IEEEauthorblockA{\textit{Researcher} \\
Senior Member, IEEE \\
Seattle, WA, USA \\
rajesh.pandey@ieee.org}
\and
\IEEEauthorblockN{Aniket Abhishek Soni}
\IEEEauthorblockA{\textit{Researcher} \\
Senior Member, IEEE \\
Brooklyn, NY, USA \\
aniketsoni@ieee.org}
}

\begin{document}
\maketitle

\begin{abstract}
While the Web has become a global platform for communication; malicious actors, including hackers and hacktivist groups, often disseminate ideological content and coordinate activities through the "Dark Web"—an obscure counterpart of the conventional web. Presently, challenges such as information overload and the fragmented nature of cyber threat data impede comprehensive profiling of these actors, thereby limiting the efficacy of predictive analyses of their online activities. Concurrently, the proliferation of internet-connected devices has surpassed the global human population, with this disparity projected to widen as the Internet of Things (IoT) expands. Technical communities are actively advancing IoT-related research to address its growing societal integration. This paper proposes a novel predictive threat intelligence framework designed to systematically collect, analyze, and visualize Dark Web data to identify malicious websites and correlate this information with potential IoT vulnerabilities. The methodology integrates automated data harvesting, analytical techniques, and visual mapping tools, while also examining vulnerabilities in IoT devices to assess exploitability. By bridging gaps in cybersecurity research, this study aims to enhance predictive threat modeling and inform policy development, thereby contributing to intelligence research initiatives focused on mitigating cyber risks in an increasingly interconnected digital ecosystem.
\end{abstract}


\section{Introduction}
The Internet has evolved into a worldwide platform that allows individuals to easily distribute, exchange, and communicate ideas. While it offers numerous benefits, the misuse of the Internet has grown increasingly problematic. Malicious entities, including hackers, extremist organizations, hate groups, and racial supremacists, are leveraging the web more than ever to carry out illegal activities, facilitate internal communications, orchestrate attacks on institutions and governments, and spread harmful ideologies. It has become commonplace to hear about cyber breaches in the news, where cybercriminals post messages targeting large corporations and organizations, demanding weapons, financial backing, and volunteers \cite{grau2016}. Consequently, there is a pressing need to gather intelligence that enhances our understanding of the tactics employed by hackers and criminal organizations.

In this paper, we focus on the Internet of Things (IoT) as a vulnerability due to its expanding presence and susceptibility to exploitation \cite{bloomberg2014}. By employing the structured methodology outlined in this study, we evaluate the degree to which cybercriminal groups target IoT systems. Through our analysis of 23 criminal websites on the Dark Web, we discovered that even seemingly harmless devices, such as small sensors, can cause substantial damage when compromised. IoT opens up an entirely new realm for cybercriminals to exploit. These hacker groups generally do not discriminate in their targets—they exploit any weakness or vulnerability they can find.

The remainder of the paper is organized as follows: Section II provides an overview of the concepts introduced in this paper; Section III outlines a methodology for collecting and analyzing Dark Web data, and identifying vulnerabilities; Section IV details how this methodology was applied in a case study of IoT using Shodan, the IoT search engine; and the final section concludes the study and explores potential solutions.

\section{Literature Review}
\subsection{Dark Web Utilization}
Hackers, criminals, and terrorists leverage the web to facilitate their illicit activities. Since the emergence of the Dark Web, numerous new marketplaces have surfaced, offering a wide and often alarming array of products, including prescription pills, meth, heroin, speed, crack, guns, stolen identities, gold, and erotica. The Dark Web also serves as a platform for spreading extremism, facilitating communication among terrorists and criminals, and sharing knowledge about illegal hacks and attacks, thereby enabling harmful activities that threaten society. Most of these markets operate using Bitcoin and other cryptocurrencies, which allow for anonymous online transactions. According to one study, criminals and terrorists use Bitcoin to sell drugs and subsequently purchase guns and ammunition. During our exploration of the dark market known as \href{http://hansamkt2rr6nfg3.onion/search/?q=hacker&c=59}{Hansa Market}, we identified a website called \href{http://armsforsd.com/index.html}{Arms and Ammunition}, which poses significant threats to nations both internally and externally. This underscores the importance of developing robust threat intelligence.

\subsection{Discovering IoT Devices Through Shodan}
We focused on IoT devices due to their growing integration into daily life and critical infrastructure. Like mobile phones, these devices are increasingly capable of learning and predicting user behavior. Despite their utility, many IoT devices operate on low-power processors and custom operating systems that often lack robust encryption capabilities, making them inherently vulnerable.

For example, a smart thermostat not only regulates temperature but also stores behavioral data on centralized servers. While a compromised thermostat in a home may seem low-risk, similar vulnerabilities in industrial control systems can lead to serious disruptions. Attackers exploiting such weaknesses could manipulate systems or target individuals based on behavioral patterns, posing threats to both personal safety and public infrastructure.

To identify exposed IoT devices, we used Shodan, a search engine that scans the internet for connected devices with open ports. Shodan can discover a wide range of systems, from webcams and servers to industrial control devices and power grids. Its effectiveness has been demonstrated in prior research, such as a study in which newly connected devices were indexed by Shodan within 14 days, revealing their IPs, ports, and system metadata \cite{bodenheim2014}.

Shodan gathers information through service banner interrogation—extracting data provided by devices in response to connection requests. With powerful filters such as IP address, port, location, and operating system, Shodan offers both attackers and defenders a valuable tool for identifying vulnerable systems and enhancing threat intelligence.

\subsection{Developing Cyber Threat Intelligence}
Our research focused on determining whether the systematic collection and processing of Dark Web content could support the development of actionable threat intelligence to protect vulnerable IoT devices. We examined how insights from underground hacker forums, marketplaces, and other illicit platforms could enhance understanding of emerging cyber threats targeting IoT ecosystems. The remainder of this paper details the methodologies we employed for data collection, the techniques used to filter and refine relevant information, and the analytical approaches applied to transform raw data into meaningful threat intelligence. By outlining these processes, we demonstrate how Dark Web intelligence can be leveraged to strengthen IoT security measures and mitigate cyber risks before they materialize into active threats.

\section{A Methodology for Collecting and Analyzing Dark Web Information}
\subsection{The Methodology}
To address the diverse threats posed by the information sources used by hackers, criminals, and terrorists, we propose a comprehensive threat intelligence framework. This framework integrates Dark Web data aggregation, machine learning-driven anomaly detection of IoT vulnerability patterns, and human-AI collaboration for real-time cyber-attack prediction.

\begin{figure*}[ht]
\centering
\includegraphics[width=\textwidth]{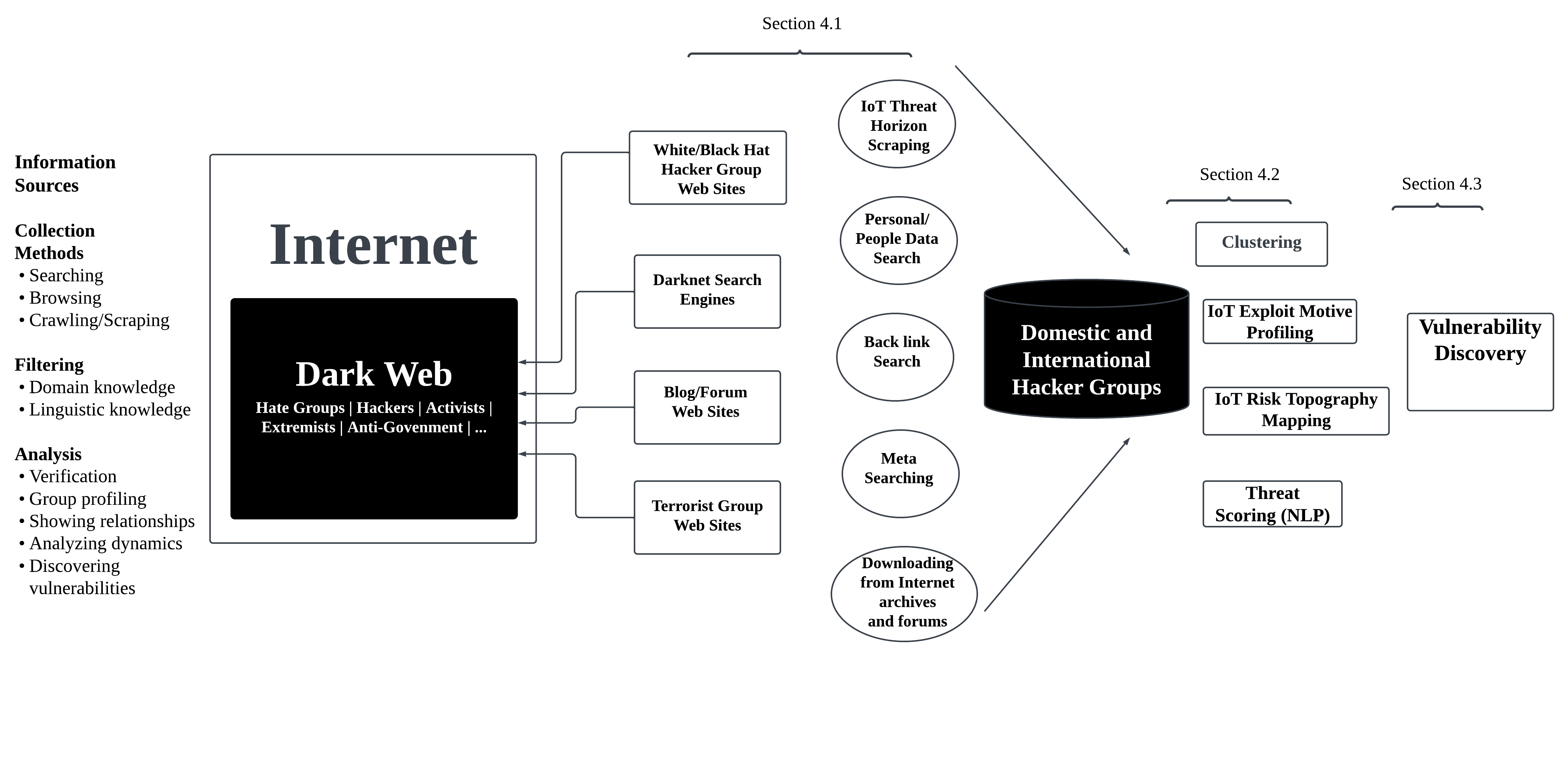}
\caption{A methodology for collecting and analyzing Dark Web information.}
\captionsetup{justification=centering}
\label{fig:methodology}
\end{figure*}

Unlike prior approaches \cite{paloalto2020}, our method correlates hacker discussions (e.g., IoT exploitation trends) with Shodan-derived device vulnerabilities to prioritize high-risk targets. Figure~\ref{fig:methodology} illustrates this methodology, designed to assist investigators in obtaining Dark Web intelligence through diverse sources, collection techniques, filtering, and analysis \cite{xu2008}.

\begin{itemize}
\item \textit{Information Sources} cover a wide range of hacker-related platforms, from public search engines to sites maintained by hacker groups, hacktivists, terrorists, or supporters—often requiring domain expertise for access.
\item \textit{Collection} employs systematic crawling starting from seed URLs or keywords, leveraging various web scraping tools to gather relevant IoT threat data.
\item \textit{Back-link Search} uses search engines like Google to trace web pages linking to hacker domains, helping uncover supporters who glorify or coordinate attacks.
\item \textit{Personal/People Data Search} through portals such as Yahoo! and MSN reveals profiles with links that expose hidden connections among hacker communities.
\item \textit{Meta-searching} aggregates results from multiple search engines using related keywords to refine and prioritize relevant information about groups like "Anonymous."
\item \textit{Filtering} requires domain and linguistic expertise to remove irrelevant data, considering the history, language, and symbolism of cyber-attackers.
\item \textit{Analysis} aids investigators by profiling IoT exploit motives, clustering hacker entities, mapping risk topographies, scoring threats based on hacker interest and exposure, and discovering vulnerabilities linked to specific targets.
\end{itemize}

\subsection{Application of the Methodology}
Although the Internet has been publicly accessible since the 1990s, the Dark Web has only emerged as a significant domain in recent years. Its anonymous and decentralized nature has facilitated a range of illicit activities, yet the development of effective methodologies for identifying vulnerabilities and tracking cybercriminal activities within this hidden space has remained a persistent challenge \cite{abbasi2007}. The absence of robust and scalable techniques has hindered law enforcement agencies, cybersecurity researchers, and policymakers in their efforts to combat cybercrimes effectively.

As discussed earlier, our proposed methodology integrates a combination of advanced data mining, web mining, and machine learning techniques while leveraging human domain expertise to refine their application. This hybrid approach enhances the ability to systematically identify and monitor criminal groups, analyze their discussions, and extract intelligence relevant to emerging cyber threats. By incorporating both automated data processing and human analytical oversight, our methodology balances computational efficiency with contextual accuracy, enabling more precise intelligence gathering. Ultimately, this semi-automated framework presents a scalable solution for uncovering hidden threat actors and understanding the evolving landscape of cybercrime on the Dark Web.

\section{Case Study: IoT Scanning to Predict Cyber Attacks}
To illustrate the effectiveness and applicability of the methodology, we applied it to the collection and analysis of web-based activities by malicious actors targeting IoT devices. These devices are especially vulnerable and projected to number 26 billion by 2030. We describe our approach in Figure~1.

\subsection{Collection}
To gather data, we first identified several suspicious URLs using web searches, referencing dark web markets. We conducted web spidering using the 'Open Web Spidering' tool. For instance, starting with keywords like “freedom fighters” and “digital Robin Hood,” we found four intriguing URLs: \href{http://hackhound.org/forums/page/index.html}{HackHound} (a forum for hackers), \href{http://anonymzn3twqpxq5.onion/read.php?2%2C8869%2C8966}{Du Bist Anonymous} (a website supporting the Anonymous group), \href{https://uniregistry.com/market/domain/cihad.net?landerid=www57aacde58c16c2.39985067}{CiHad.net} (a suspicious website associated with a terror group), and \href{http://rrcc5uuudhh4oz3c.onion/}{Intel Exchange} (a political ideology website). Using these categories as references, we uncovered additional websites during the data collection process.

We then performed the same searches on dark web search engines such as Torch, Grams, and Ahmia. Although the results were scattered, we stored them nonetheless, as they primarily pointed to dark web markets. Manually browsing these markets revealed offers of both hacking services and services required for hacking. We selected keywords such as “hacking services,” “free world,” “personal army,” and “Internet of Things,” which yielded a variety of relevant results. Overall, we collected information from hundreds of sources, and a promising future direction would be to study how different data collection techniques affect the quality and depth of subsequent analysis.

\subsection{Filtering}
We conducted two rounds of filtering to refine the dataset. In the first round, we manually excluded websites that were clearly unrelated to cybersecurity or hacking, such as news portals, gaming platforms, and adult content sites. We retained websites hosting forums, technical blogs, ideological discussions, and dark market listings offering hacking tools, data leaks, or illicit services such as ammunition. This process reduced the dataset to thirty-two websites.

In the second round, we refined the selection further. Given limitations in language coverage and domain-specific nuance, we manually removed sites that appeared redundant, linguistically inaccessible, or thematically distant from our focus. To support this process, we applied natural language processing (NLP) and basic machine learning (ML) techniques to tag security-relevant keywords and compute a threat relevance score for each website. This score was generated using TF-IDF-based feature extraction and a logistic regression model trained on a small labeled dataset comprising known malicious and benign cybersecurity forums. Websites with higher scores—indicating stronger alignment with cyber threat indicators—were prioritized for deeper analysis. This hybrid manual-automated filtering approach helped reduce subjectivity and improved consistency in identifying high-risk content.

\subsection{Analysis}
\subsubsection{Classification and Usage of Websites}
We conducted two rounds of filtering. First, we manually removed unrelated sites, such as news, games, and porn sites, which had no reference to hacking. We retained websites of forums, blogs, ideological discussions, and dark markets for hacking and ammunition services. After this round, thirty-two sites remained.

Second, given our limited domain expertise and knowledge of languages, we initially manually removed some websites that felt less relevant and were similar to English forum websites in our results. Subsequently, we used natural language processing (NLP) and machine learning (ML) for keyword tagging and threat scoring.

We performed clustering and classification of the remaining 23 websites. These websites provided a diverse range of information necessary for our analysis. Table~\ref{tab:darkweb_sites} lists these websites with hyperlinked URLs and descriptions.

\begin{table}[ht]
    \centering
    \begin{tabular}{|p{3cm}|p{4cm}|}
        \hline
        \rowcolor{black}
        \textcolor{white}{\textbf{Website}} & \textcolor{white}{\textbf{Description}} \\
        \hline
        \rowcolor{gray!30}
        \href{http://hansamkt2rr6nfg3.onion/search/?q=hacker&amp;c=59}{Hansa Market} & Dark market - Hacking Services \\
        \hline
        \rowcolor{gray!30}
        \href{http://armsforsd.com/index.html}{Arms and Ammunition} & Guns \& Ammunition \\
        \hline
        \rowcolor{gray!30}
        \href{http://anonymzn3twqpxq5.onion/read.php?2%2C8869%2C8966}{Du Bist Anonymous} & Hacking services (Anonymous Hackers Site) \\
        \hline
        \rowcolor{gray!30}
        \href{http://2ogmrlfzdthnwkez.onion/info.php}{Rent A Hacker} & Hacking services \\
        \hline
        \rowcolor{gray!30}
        \href{http://opnju4nyz7wbypme.onion/weblog/index.html}{Alpha7-Bravo- Blog} & Political Hate Ideology \\
        \hline
        \rowcolor{gray!30}
        \href{http://duskgytldkxiuqc6.onion/comsense.html}{Thoman Paine Common Sense} & Political Hate Ideology \\
        \hline
        \rowcolor{gray!30}
        \href{http://oxwugzccvk3dk6tj.onion/uk/res/16.html}{Random United Kingdom} & Hate group forum \\
        \hline
        \rowcolor{gray!30}
        \href{http://rrcc5uuudhh4oz3c.onion/}{Intel Exchange} & Hacking Forum \\
        \hline
        \rowcolor{gray!30}
        \href{http://6sgjmi53igmg7fm7.onion/index.php?title=Main_Page}{Bugged Planet} & Hate group forum \\
        \hline
        \rowcolor{gray!30}
        \href{http://f2mz6ttcwyslnz5u.onion/}{Jaggers Blog} & Hacking Blog and Forum \\
        \hline
        \rowcolor{gray!30}
        \href{http://zw3crggtadila2sg.onion/imageboard/}{TorChan} & Hacking Discussions Forum \\
        \hline
        \rowcolor{gray!30}
        \href{https://app.hackerwebapp.com/}{HackerWeb} & Cyber Security Hacking Discussion Forum \\
        \hline
        \rowcolor{gray!30}
        \href{http://hackhound.org/forums/page/index.html}{HackHound} & Cyber Security Hacking Discussion Forum \\
        \hline
        \rowcolor{gray!30}
        \href{https://hackerstribe.com/}{HackersTribe} & Hacking Discussions Forum \\
        \hline
        \rowcolor{gray!30}
        \href{http://www.school-of-hack.net/}{School-of-HackNet} & Hacking Discussions Forum \\
        \hline
        \rowcolor{gray!30}
        \href{http://www.ic0de.org/activity.php?s=e72472d21ab6ba2672b1b13f8c012671}{iC0de} & Cyber Security Hacking Discussion Forum \\
        \hline
        \rowcolor{gray!30}
        \href{http://valhallaxmn3fydu.onion/products/25348}{Valhalla} & Dark market - Hacking Services \\
        \hline
        \rowcolor{gray!30}
        \href{https://rhyliv.com/}{Rhyliv} & Cyber Security Hacking Discussion Forum \\
        \hline
        \rowcolor{gray!30}
        \href{http://www.alokab.com/}{Alokbab} & Suspicious Terror Group Site \\
        \hline
        \rowcolor{gray!30}
        \href{http://www.fadaian.org/file/manch.uk.html}{Fadaian} & Suspicious Terror Group Site \\
        \hline
        \rowcolor{gray!30}
        \href{https://uniregistry.com/market/domain/cihad.net?landerid=www57aacde58c16c2.39985067}{CiHad.net} & Bogus Site – Suspicious Terror Group \\
        \hline
        \rowcolor{gray!30}
        \href{http://oasisnvwltxvmqqz.onion/79}{Oasis} & Dark market - Hacking Services \\
        \hline
    \end{tabular}
    \caption{Websites collected with the implementation of the methodology.}
    \label{tab:darkweb_sites}
\end{table}

\begin{figure}[htbp]
    \centering
    \includegraphics[width=\linewidth]{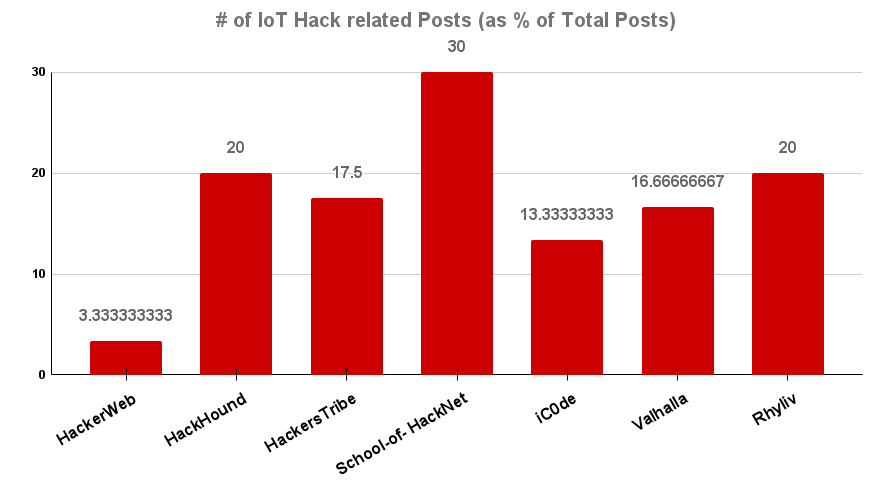}
    \caption{Statistics of IoT device hacking discussions within December 2023 - July 2024.}
    \label{fig:iot_hacking_stats}
\end{figure}

We followed the following steps to analyze “IoT Hacking” or “hacking IoT devices” discussions on some of these websites:

\begin{itemize}
    \item We filtered cybersecurity hacking discussion forums from the list above, ignoring the rest of the websites.
    \item We manually browsed through these forums and searched for the keywords listed.
    \item We gathered statistics as recent as possible (up to 2024) and generated a graph (Figure~\ref{fig:iot_hacking_stats}) and a table (Table~\ref{tab:iot_hacking}).
\end{itemize}

The graph in Figure~\ref{fig:iot_hacking_stats} displays the percentage of IoT hacking-related discussions occurring in the seven forums we selected for analysis. The x-axis shows the names of these forums, while the y-axis shows the percentage of IoT hacking discussions. The "Hacker Web" forum, which aggregates 18 major forums, has 3.3\% of its posts discussing IoT hacking. Although this percentage is lower, it does not imply fewer discussions, given the forum's size. The other forums show 12\% to 30\% of their topics discussing IoT hacking.

\begin{table}[htbp]
    \centering
    \renewcommand{\arraystretch}{1.3}
    \begin{tabular}{|p{2cm}|p{1cm}|p{1cm}|c|}
        \hline
        \rowcolor{black}
        \textcolor{white}{\textbf{Website}} & \textcolor{white}{\textbf{Searched}} & \textcolor{white}{\textbf{Found}} & \textcolor{white}{\textbf{\# of Posts}} \\
        \hline
        \rowcolor{gray!30}
        \href{http://hackhound.org/forums/page/index.html}{HackHound} & Internet of Things & Botnet, Malware, Rats for Android devices & 4 \\
        \hline
        \rowcolor{gray!30}
        \href{https://hackerstribe.com/}{Hackers Tribe} & Hack Internet of Things & Sensors, File Sharing in devices & 5 \\
        \hline
        \rowcolor{gray!30}
        \href{http://www.school-of-hack.net/}{School-of-HackNet} & Hack devices & Phones Listening to Network & 1 \\
        \hline
        \rowcolor{gray!30}
        \href{https://app.hackerwebapp.com/}{HackerWeb} & Hack Internet of Things & Knife for IoT: XMPP & 1 \\
        \hline
    \end{tabular}
    \caption{This table shows the keywords searched and data found on the mentioned websites.}
    \label{tab:iot_hacking}
\end{table}

By manually browsing through these forums, we analyzed the most frequently discussed topics related to IoT hacking. We selected four websites listed in Table~\ref{tab:iot_hacking}, searched for the keywords in the 'Searched' column, and recorded the results and total post counts in the 'Found' and '\# of Posts' columns, respectively. Our analysis revealed that botnets, malware, and sensors are the most discussed methods for exploiting IoT devices. To determine how easily these vulnerabilities can be exploited, we used the IoT search engine, Shodan.

\subsubsection{Shodan: Discovering Vulnerable IoT Devices}
Using Shodan, we followed these steps for our analysis:

Upon executing the Python script detailed in Listing~\ref{code:shodan}, we obtained 582 results for sensors. In the try block of the code, we captured the number of results and filtered information such as company name, IP address, port, operating system, and data details, including content type, authentication protocol, and location specifics like country and city. The results from our Shodan search provided comprehensive information about the devices, including company name, location, IP address, and port. This level of detail facilitates various forms of analysis.

\begin{enumerate}
    \item Configured Shodan’s REST API with Python.
    \item Searched for the keyword \texttt{sensor}.
    \item Exported the results in \texttt{.csv} format.
    \item Used Excel for further analysis.
\end{enumerate}
\vspace{-2mm}
\begin{lstlisting}[language=Python,caption={Python code used for searching in Shodan and saving results to CSV},label=code:shodan]
# Importing the necessary libraries
import shodan
import csv

# Define your Shodan API key
SHODAN_API_KEY = "your_api_key"

# Initialize the Shodan API client
api = shodan.Shodan(SHODAN_API_KEY)

# Open a CSV file to store the results
with open('shodan_results.csv', 'w', newline='') as csvfile:
    fieldnames = ['IP', 'Data']
    writer = csv.DictWriter(csvfile, fieldnames=fieldnames)

    # Write the header to the CSV file
    writer.writeheader()

    try:
        # Perform a search query using the Shodan API
        results = api.search('sensor')
        
        # Print the total number of results found
        print('Results found: {}'.format(results['total']))
        
        # Loop through each result and save it to the CSV
        for result in results['matches']:
            print('IP: {}'.format(result['ip_str']))
            print(result['data'])
            
            # Write each result to the CSV file
            writer.writerow({'IP': result['ip_str'], 'Data': result['data']})

    # Handle API-related errors
    except shodan.APIError as e:
        print('Error: {}'.format(e))
\end{lstlisting}
\begin{figure}[H]
    \centering

    \begin{subfigure}[b]{0.45\linewidth}
        \centering
        \includegraphics[width=\linewidth]{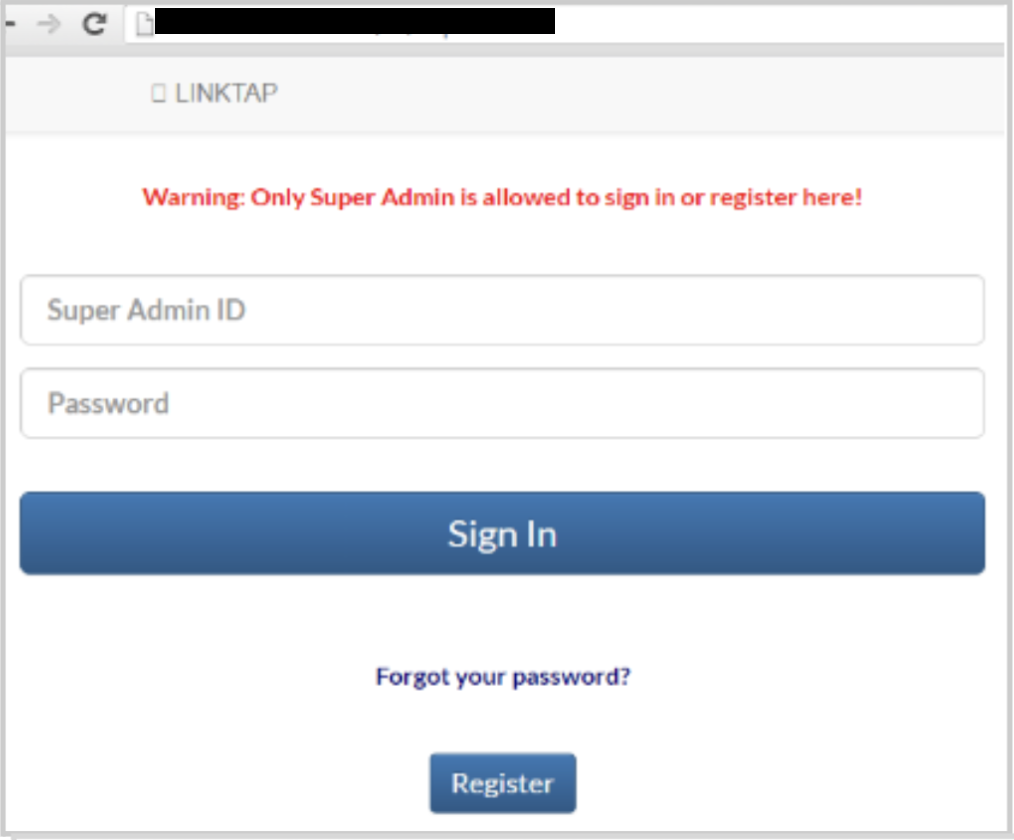}
        \caption{Temperature-sensing device}
        \label{fig:sensor_img3}
    \end{subfigure}
    \hfill
    \begin{subfigure}[b]{0.45\linewidth}
        \centering
        \includegraphics[width=\linewidth]{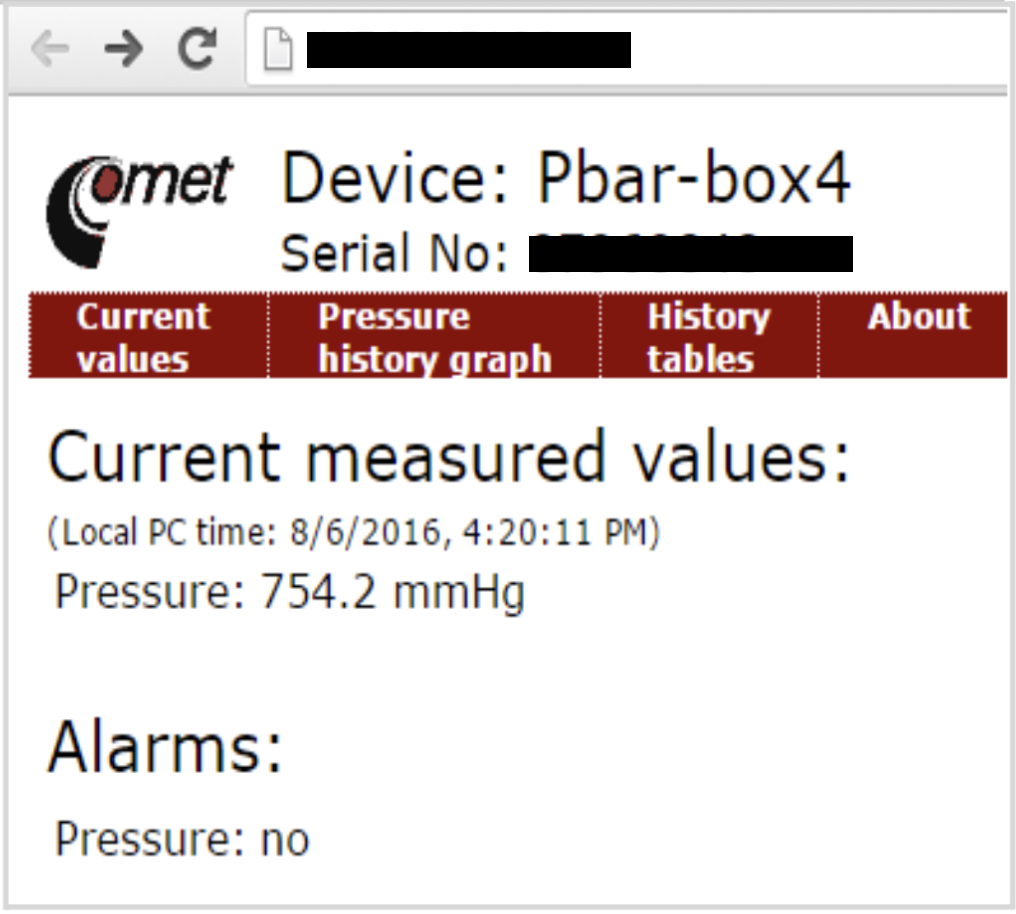}
        \caption{Other sensor devices}
        \label{fig:sensor_img4}
    \end{subfigure}

    \vspace{2mm}

    \begin{subfigure}[b]{0.45\linewidth}
        \centering
        \includegraphics[width=\linewidth]{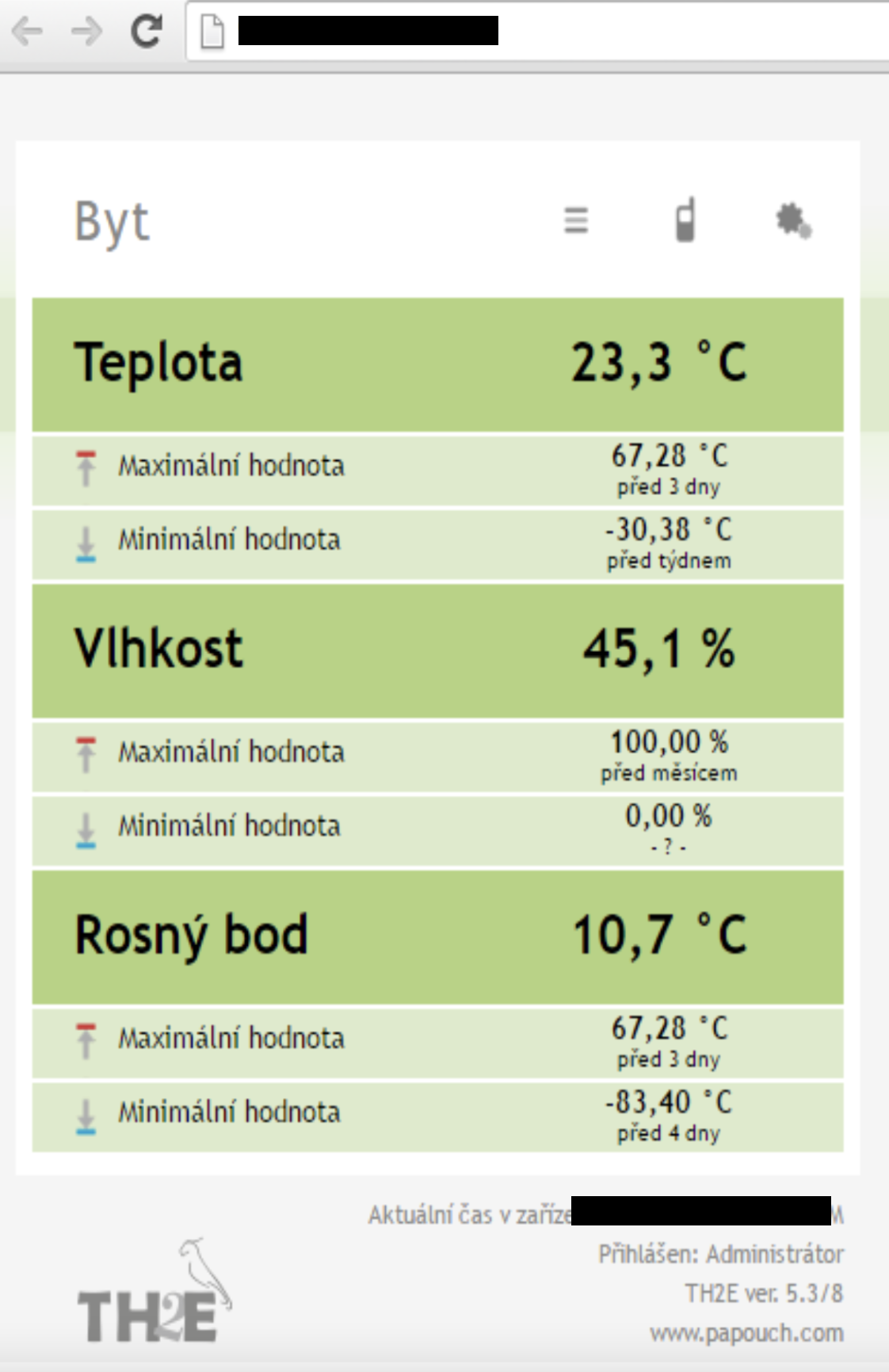}
        \caption{Pressure-sensor device}
        \label{fig:sensor_img2}
    \end{subfigure}
    \hfill
    \begin{subfigure}[b]{0.45\linewidth}
        \centering
        \includegraphics[width=\linewidth]{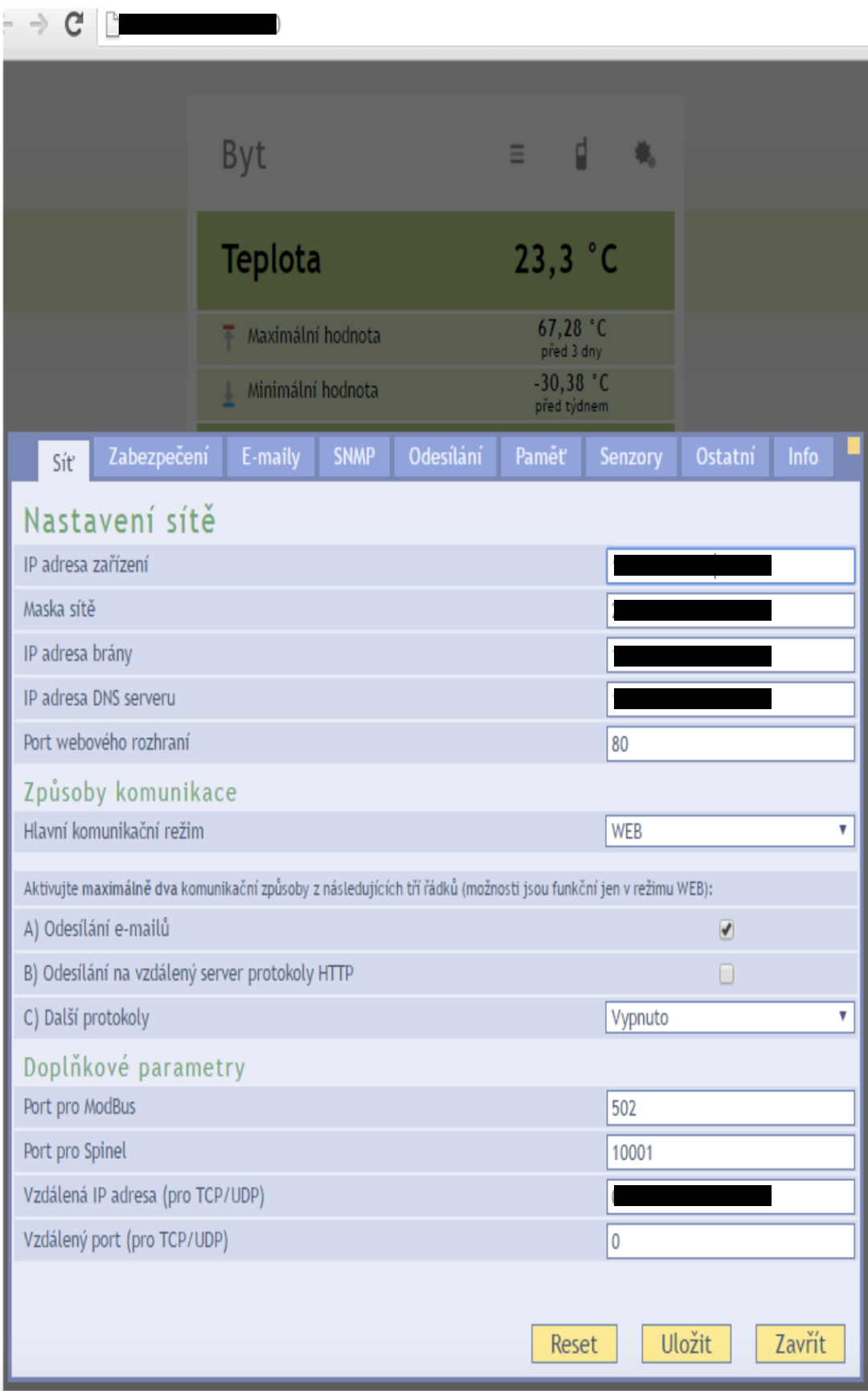}
        \caption{Admin control panel}
        \label{fig:sensor_img1}
    \end{subfigure}

    \caption{Examples of vulnerable IoT devices discovered via Shodan}
    \label{fig:all_sensors}
\end{figure}

First, we filtered the results to include only those with port 8080 and directly accessed some of the IP addresses. As shown in Figure~\ref{fig:sensor_img3}, we encountered a page requesting login credentials. This is concerning because it exposes the device to a range of security vulnerabilities, such as cross-site scripting, denial-of-service attacks, SQL injection, unexpected application termination, and arbitrary code execution. Additionally, the connection was not secured over SSL. 

Second, we manually examined the IP addresses providing direct access. As illustrated in Figures~\ref{fig:sensor_img4} and \ref{fig:sensor_img2}, we gained access to a pressure-sensing device named 'Pbar-box4' and a temperature-sensing device named 'Byt.' We had full control over these devices, allowing us to view past temperature readings and manipulate administrative settings, including changing email addresses where data was being sent. Figure~\ref{fig:sensor_img1} demonstrates an example of the control panel for the 'Byt' device.

\section{Conclusions}
Analyzing and collecting Dark Web information is exceedingly challenging due to cyber-attackers' ability to conceal their identities and erase traces of their activities online. The sheer volume of web information complicates the task of obtaining a comprehensive understanding of their operations, making it even more difficult to predict their next targets. In this article, we propose a methodology to address these challenges. By employing web mining, content analysis, and visualization techniques, the methodology leverages various information sources and analyzes 23 hacker websites for IoT vulnerabilities. Information clustering is used to differentiate between novice and experienced hackers on the web. Our detailed evaluation indicates that the methodology produced promising results, which could significantly aid in discovering new vulnerabilities and their sources.

For IoT, we propose the following solutions: 
\begin{enumerate}
\item Strict security standards must be established for the manufacturing of IoT hardware and software.
\item IoT software should be developed by dedicated companies that rigorously adhere to IoT security protocols.
\item Best practices must be followed by users and those configuring IoT devices.
\end{enumerate}

This research was conducted from multiple angles, and we believe it will contribute to the existing literature on cyber-security, potentially guiding policy-making and threat intelligence research.

\section{Declaration}
\subsection*{Data Availability}
The datasets generated and analyzed during this study are available from the authors upon reasonable request. Figures 1 (methodology flowchart), 2 (forum discussion statistics), and 4 (anonymized IoT device screenshots) are included in the manuscript, with all sensitive information redacted to prevent misuse.

\subsection*{Funding}
The authors received no specific funding for this work.

\bibliographystyle{IEEEtran}

\end{document}